\documentclass[twocolumn,aps,showpacs,preprintnumbers,superscriptaddress]{revtex4-1}
\usepackage{graphicx}
\usepackage{dcolumn}
\usepackage{revsymb4-1}
\usepackage{amsmath,amssymb}
\usepackage{bm}

\def\ao{ Appl.\  Opt.\ }
\def\aop{ Adv.\ Opt.\ Photon.\ }

\def\josaa{ J.\ Opt.\ Soc.\ Am.\ A }
\def\josab{ J.\ Opt.\ Soc.\ Am.\ B }

\def\oc{ Opt.\ Commun.\ }
\def\ol{ Opt.\ Lett.\ }

\def\prl{ Phys.\ Rev.\ Lett.\ }

\begin{document}
\preprint{}
\title{Generation of optical vortices in a degenerate optical resonator with an intra-cavity spiral phase plate}
\author{Yuan-Yao Lin$^\ast$}
\affiliation{Department of Photonics, National Sun Yat-sen University, Kaohsiung, Taiwan, R.O.C. 804}
\email{yuyalin@mail.nsysu.edu.tw}
\author{Chia-Chi Yeh}
\affiliation{Department of Photonics, National Sun Yat-sen University, Kaohsiung, Taiwan, R.O.C. 804}
\author{Hsien-Che Lee}
\affiliation{Department of Photonics, National Sun Yat-sen University, Kaohsiung, Taiwan, R.O.C. 804}

\begin{abstract}
\noindent 
We propose to generate optical vortices using a degenerate optical resonator with an intra-cavity spiral phase plate (SPP). The rays retracing skewed V-shaped paths in the resonator are phase-locked to form vortex laser with wavefront dislocation mirroring the topological charge of the SPP. Experimental demonstrations on diode-pumped solid state system using Nd:YAG crystal emitted randomly polarized optical vortices at a wavelength of 1064nm when the pump power went above 1.52W and the slope efficiency was 0.19. For long-term operations a power fluctuation of 2.2\% and pointing stability of 2.6$\mu$rads was measured. This system also serves an useful platform to study the laser dynamics and to combine radiations coherently.
\end{abstract}
\maketitle
\noindent Optical vortex (OV) is more than a beam of donut-shaped intensity profile. It carries well-defined orbital angular momentum (OAM) in the photons within~\cite{allen2003} which distinguishes OV beams from vector beams of radail or azimuthal polarizations~\cite{Mushiake1972, AO.18.581}. The unique property of OV beam have attracted growing attentions due to a wide range of promising applications including microscopy~\cite{microscope2016}, particle manipulation~\cite{allen2003}, astronomy and cosmology~\cite{hawking_rad} and sub-Peta hertz bit-rate optical communications~\cite{Tbit}. OAM further provides additional information for wavefront reconstruction and thus can enhance the channel capacity in free space communications under the effect of atmospheric turbulence~\cite{srep.7.43233}. Although a number of techniques were devised to generate OV with helical wavefront from fundamental gaussian beams by exploiting external resonator conversion element such as spiral phase plate (SPP)~\cite{aop.3.161}, holograms~\cite{aop.3.161} that modulated the wavefronts or Q-plate that induced spin-orbital couplings~\cite{JO.13.064001}, not one of these approaches achieved a satisfactory conversion efficiency and beam quality. 

Direct OV beam lased from resonator is more desirable because of better beam quality in the sense of coherence and mode purity. Paraxial beams emitted from low-loss optical cavity such as Laguerre-Gaussian (LG) mode can carry OAM for any radial index provided a non-vanishing azimuthal index. Although in a Bessel-Gauss resonator high quality LG beams were generated~\cite{JOSAA.20.2113}, the handedness of OAM in the beam is not determined unless azimuthal symmetry-breaking (ASB) mechanism was presented in the laser resonators.  By introducing intra-cavity spiral phase elements in laser resonators, LG modes, could be discriminated~\cite{OC.169.115} and lead to the production of pure helical laser beam~\cite{OC.182.205}.  Replacing spiral phase element with a pair of Q-plates, higher-order Poincar\'{e} sphere beams was also generated from a laser resonator~\cite{nphoton.10.327}. Recently LG beams of zero radial index and unit azimuthal index with controlled handedness are generated when using an annular-shaped pump beam and an intra-cavity mode selection element to produce selective loss to the standing waves of two distinct handednesses~\cite{ol-30-3903}. Moreover inserting a tilted Fabry-Perot etalon into a laser resonator differentiated the Poynting vectors of the LG modes of opposite helical wavefront structure~\cite{OL.40.399}. Shaping pump region inside the gain media, the spatial overlap of modes with varying azimuthal index could be adjusted and thus modes were selected in both continuous wave (CW)~\cite{OC.296.109, JOSAA.27.2072} and active Q-switch regime~\cite{OC.347.59}.

Degenerate optical cavities have been studied for more than four decades. They support rays that retrace their paths after one or more than one round trips when off-axis ray is taken as the optical axis~\cite{AO.8.189}. Conventional low-loss optical resonators formed by spherical mirrors can be designed to be a degenerate linear system in which the rays can complete ecliptic or nonecliptic closed path of planar or non-planar categories~\cite{AO.9.385}. Notably the multipass transverse modes (MPT) regarding to the "N-fold" degeneracy are aberration-limited and are not any superposition of the single-path transverse (SPT) modes such as Hermite-Gaussian and LG modes~\cite{JOSAB.18.7,LPH.25.023001}. MPT modes were realized in several laser systems including the CO$_2$ gas laser by careful adjustment and alignment of the cavity mirrors~\cite{AO.9.385} and also in diode pumped solid-state laser (DPSS) using an off-axis pumping geometry and  a fractionally degenerate resonator design~\cite{JOSAB.18.7}. With the capability to support more-or-less arbitrary transverse beam pattern, degenerate resonators serve as a great platform to apply for optimizing the extraction efficiency of solid-state laser~\cite{OL.19.1134}, the manipulation of spatial coherences of a laser system~\cite{OL.38.3858} and intra-cavity synthesis of various laser waveforms~\cite{PAO.3.757}.

\begin{figure}[h]
\includegraphics[width=8cm]{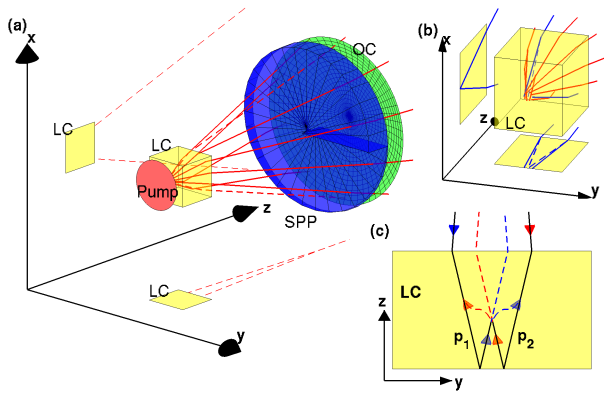}
\caption{\label{fig:1}(Color online) (a) Schematics of the vortex resonator configuration. Pump is pumping laser; LC is the laser gain crystal end coated with HR at resonating wavelength; SPP is the spiral phase plate; OC is the output coupler. The lines and dashed lines in red stand for the ray trajectories retracing in the cavity. The projections of the ray trajectories onto the $y-z$ plane and $x-z$ plane are plotted respectively to show the skewed ray trajectory presented by the dashed line in Fig.~\ref{fig:1}(a). Ray trajectories in the laser crystal and the projections on $y-z$ and $x-z$ planes are shown in the blow-up figure (b). (c) delineates the formation of closed orbital through composite interactions. The solid lines are the ray trajectories and the dashed lines are the virtual path describing the composite interactions.}
\end{figure}

To generate high quality continuous wave OV beams at a desired wavelength, we propose a vortex resonator, as shown in Fig.~\ref{fig:1}(a), using a near-semi-spherical configuration in which a transmissive SPP is placed near a curved OC mirror. The gain media has a planar end surface that is high-reflection (HR) coated at the resonating wavelength to serve as a cavity mirror. Notably the vortex resonator proposed in Fig.~\ref{fig:1}(a) does not support LG modes although it is similar to the configuration proposed by Oron {\it et at.}~\cite{OC.182.205} that supported and discriminated LG modes at the first sight. Without a cylindrical lens placed inside the cavity to break the mirror symmetry, LG modes travelling in the vortex resonator illustrated in Fig.~\ref{fig:1}(a) experiences additions of twice the topological charges of the SPP for every round trip. The circulating LG modes are then lost from the vortex resonator because their beam radii exceed the dimension of the cavity mirrors as the carrying topological charges increase. It can be inferred that any SPT modes are not permitted to oscillate in the vortex resonators as a consequence of the completeness of the orthogonal LG modes.

In terms of ray optics, the vortex resonator in Fig.~\ref{fig:1}(a) is a near semi-spherical resonator of 2-fold degeneracy perturbed by SPP.  Without loss of generality, it is delineative to examine the trajectory illustrated by dashed line of Fig.~\ref{fig:1}(a) along with its projections onto $x-z$ and $y-z$ planes. On the $y-z$ projection, the $y-$ component of the ray is refracted by the SPP~\cite{josaa.30.2526} and is vanished at the OC and on the $x-z$ projection the $x-$ component is normal to the reflecting surface of OC. The ray thus can trace back directly to the laser crystal. In the laser crystal, as shown in Fig.~\ref{fig:1}(b), the $x-$ and $y-$ components of the ray are further focused by thermal lens and results in two spots at the HR coated end surface of the laser crystal, which is illustrated in blue lines of Fig.~\ref{fig:1}(b). Notice that the specific pair of trajectories is connected when projected on the $x-z$ plane but are not connected on the $y-z$ projection. By rotation symmetry, the ray trajectories form a ring of small radius on the HR-coated end surface of the laser crystal illustrated in Fig.~\ref{fig:1}(b). To have stable radiation field oscillating in vortex resonator, the ray trajectories must retrace a closed orbital. In the proposed resonator configuration, the closed orbitals can be established by composite beam interactions in the laser crystal including the nonlinear interactions and linear coupling by scattering and diffractions. As illustrated in Fig.~\ref{fig:1}(c), radiation field tracing back to laser crystal by the beam path $p_1$ along the direction marked by arrow in blue is reflected by the HR-coated end surface of the laser crystal. It is amplified by stimulated emission in laser crystal and at the same time coupled to beam path $p_2$ by composite interactions. The radiation field along $p_2$ leaving laser crystal can be retraced back to the laser crystal following the direction marked by arrow in red. It is then coupled to the beam path $p_1$, reflected by OC and returns to laser crystal so that a closed orbital is completed. It takes two round trips in the resonator to complete a closed orbital and it is thus a MPT mode that oscillates. Notably the composite interaction could be strong when beam paths $p_1$ and $p_2$ are close to each other at the HR-coated end surface of the laser crystal. By rotation symmetry, the vortex resonator supports a cone of closed skewed V-shaped orbitals and each of which is an independent laser beam radiated from the resonator. Although some particular closed orbital can be formed given sufficient astigmatism in thermal lens~\cite{QE.26.51} as illustrated by the dashed lines in blue of Fig.~\ref{fig:1}(b), yet astigmatism alone cannot support the beam paths at all azimuthal angles in the radiation cone.

The OV beam created in the proposed vortex resonator is the superposition of phase-locked MPT modes instead of LG modes. In the laser crystal, laser beams in the radiation cone following the closed skewed V-shaped orbitals overlap with the neighboring beams and strong couplings occur. The mutual coherence among the laser beams is established and the laser beams become in-phase to achieve the minimum system energy analogous to the coupled oscillator system described by Kuramoto model~\cite{PRL.110.184102}. The OAM of the radiation field as a whole simply emulates the topological charge of the SPP in the resonator. Notably the phase singularity induced by the SPP hinders the on-axis (z-axis) rays to resonate and the radius of the radiation cone in this vortex resonator is mostly determined by the aberrations of the resonator system~\cite{AO.9.385}, namely, the SPP~\cite{josaa.30.2526} and thermal lensing~\cite{QE.26.51}.  

\begin{figure}[h]
\includegraphics[width=8cm]{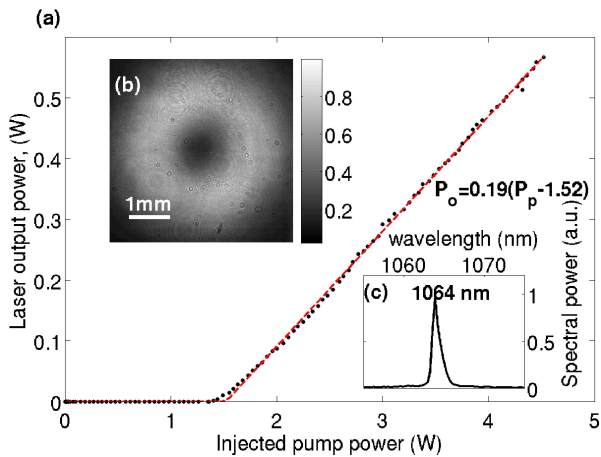}
\caption{\label{fig:2} (Color online) (a) Measured output power of OV laser as a function of injected pump power. The dots in black and dashed line in red are the experiment data and the fitted curve, respectively. (b) plots the donut-shaped intensity profile captured by a CMOS camera placed near OC. The spectrum recorded in (e) is measured by a miniature spectrometer.}
\end{figure}

\begin{figure}[h]
\includegraphics[width=8.3cm]{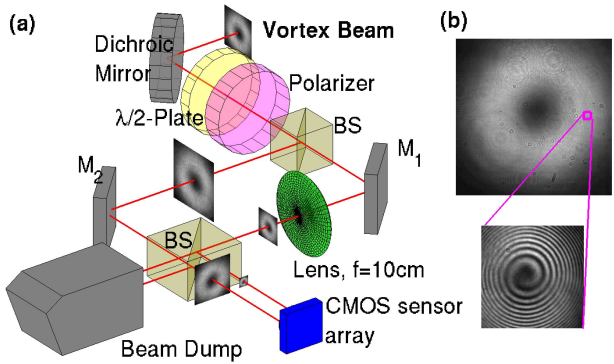}
\caption{\label{fig:3} (a) Schematics of a shifted Mach-Zehnder interferometer. $M_1$ and $M_2$ are reflecting mirrors. Two BSs are non-polarizing cube beam splitters. $\lambda/2$-Plate is the half wave plate. CMOS sensor array captures the image of combined beams. The blow-up (b) shows the interference fringe of a spiral phase.}
\end{figure}
Based on the vortex resonator lay out in Fig.~\ref{fig:1}(a), a DPSS laser system was built that emitted OV beam at a wavelength of 1064nm. The resonator length was about 19.5cm starting from the OC of which the surface has a radius of curvature of 20cm, to the planar HR coated end surface of a Nd:YAG crystal. The OC has a 10\% transmission at a wavelength of 1064nm and high-transmission at a wavelength of 808nm. A commercially available uncoated SPP made of polymer replicated on a glass substrate was located in the resonator close to the OC. Nd:YAG crystal was chosen as the gain media because its isotropic property simplifies the vortex generation scenario. The Nd:YAG crystal was pumped by diode laser at the wavelength of 808nm which was focused by a lens of focal length of 50mm to yield a focal spot size of 324$\mu$m in the laser gain medium.  The output power from the laser was measured after a dichroic mirror to remove the residue pump power. It is shown in Fig.~\ref{fig:2}(a) that when the pump power injected to the laser crystal went beyond a fitted value of 1.52W, it started to lase a randomly polarized coherent radiation at a slope rate of 0.19W per watt pumped.  The donut-shaped intensity distribution was able to be observed by an infrared detector cards directly and was captured by a CMOS camera as pictured in Fig.~\ref{fig:2}(b) which was recorded at a pump power of 3.5W. The donut-shaped beam had a beam radius of 750$\mu$m at the OC and is diverging at an angle of 4mrads. At the HR coated end surface of the laser crystal, the radius of the radiation cone was estimated to be shorter than 75$\mu$m.  The spectrum intensity measured by a miniature spectral meter plotted in Fig.~\ref{fig:2}(c) showed that the central wavelength of the OV was 1064nm. Pumped by a diode laser of power fluctuation less than 2\%, long-term stable operation was recorded to reveal a power fluctuation as low as 2.2\% and a pointing fluctuation as small as 2.6$\mu$rads. The wavefront structure of the radiation field was investigated by a self-built Mach-Zehnder interferometer as set up in Fig.~\ref{fig:3}(a).  A non-polarizing cube split the beam into two. The one reflected by mirror, $M_1$, was focused by a lens of 10mm focal length and its center was deliberately shifted. The focused beam interfered with a diverging beam reflected by $M_2$. On the CMOS sensor array camera, the focused beam had a beam radius much smaller than the diameter of the ring cross section and covered only on a small portion of the diverging beam that served as a plane wave reference. A clear and stable helical phase pattern shown in Fig.~\ref{fig:3}(b) was observed to expose a topological charge of $+1$ that emulated the intra-cavity SPP. Notably the phase structure of the generated randomly polarized OV beam displayed no polarization dependence which is inspected by equipping a rotatable half-wave plate and fixed polariser in front of the interferometer. 

\begin{figure}[h]
\includegraphics[width=8.3cm]{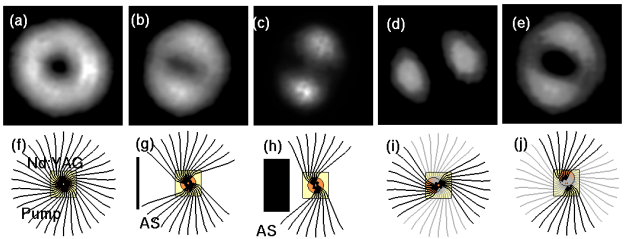}
\caption{\label{fig:4} (Color online) (a-c) recorded series of spatial intensity distribution of the laser beam when an obstacle gradually moved into the vortex resonator in horizontal direction shown in (f-h), respectively. Intensity profile in (d) and (e) were recorded when the pump beam translated along directions plotted in (i) and (j), respectively. The lines in (f-j) showed the allowed orbitals of the MPT modes schematically. The gray lines in (i-j) were those orbitals that faded away under the gain competition.}
\end{figure}

By moving an aperture stop gradually into the resonator in horizontal directions, as illustrated in Fig.~\ref{fig:4}(f-h), donut-shaped intensity distribution were expected to turn ellipse and dipole sequentially in a symmetric way depicted before laser output ultimately faded away. In Fig.~\ref{fig:4}(a-c), the recorded images agreed with the prediction and justified that these MPT modes followed skewed V-shaped close orbitals and their phases were locked to form the OV. Furthermore by translating the transverse location of the pump beam along horizontal and vertical directions in the laser crystal illustrated in Fig.~\ref{fig:4}(i) and Fig.~\ref{fig:4}(j), the superposition of the MPT modes can be tailored to exhibit dipole patterns presented in Fig.~\ref{fig:4}(d) and Fig.~\ref{fig:4}(e), respectively as a result of gain competition in a homogeneously broadened gain media such Nd:YAG crystal. It suggested This system also serve an useful platform to study the laser dynamics and to combine radiations coherently.

To conclude, we proposed that a degenerate optical resonator with an intra-cavity spiral phase plate (SPP) enables OV beam to lase as a superposition of MPT modes retracing a skewed V-shaped paths. Radiation fields along these paths complete closed orbitals by the aid composite interactions in the laser crystal and are also phase-locked to form vortex laser with wave front dislocation agreeing to the topological charge of the SPP. When realized, OVs were generated and the skewed V-shaped light paths were justified. The laser operated with a threshold pump power of 1.52W and a slope efficiency of 0.19W per Watt pumped. The OV beam generated by this resonator was stable within power fluctuation of 2.2\% and pointing deviation of 2.6$\mu$rads. This work not merely provides an efficient and robust approach that generates high quality OV beam but also suggests a platform to study the subtle dynamics of nonlinear coupled system.
\section*{Funding Information}
\noindent Ministry of Science and Technology (MOST) (MOST-104-2112-M-110-001, MOST-105-2112-M-110-002).

\end{document}